# Enhanced Circuit Densities in Epitaxially Defined FinFETs (EDFinFETs) over FinFETs

S. Mittal, A. Nainani, *Member, IEEE*, M.C. Abraham, *Member, IEEE*, S. Lodha, *Member, IEEE* and U. Ganguly, *Member, IEEE*

*Abstract*—FinFET technology is prone to suffer from Line Edge Roughness (LER) based $V_T$ variation with scaling. To address this, we proposed an <u>E</u>pitaxially <u>D</u>efined (ED) FinFET (EDFinFET) as an alternate to FinFET architecture for 10 nm node and beyond. We showed by statistical simulations that EDFinFET reduces LER based $V_T$ variability by 90% and overall variability by 59%. However, EDFinFET consists of wider fins as the fin widths are not constrained by electrostatics and variability (cf. FinFETs have fin width ~ $L_G$/3 where $L_G$ is gate-length). This indicates that EDFinFET based circuits may be less dense. In this study we show that wide fins enable taller fin heights. The ability to engineer multiple STI levels on tall fins enables different transistor widths (i.e. various W/Ls e.g. 1-10) in a single fin. This capability ensures that even though individual EDFinFET devices have ~2× larger footprints than FinFETs, EDFinFET may produce equal or higher circuit density for basic building blocks like inverters or NAND gates for W/Ls of 2 and higher.

*Index Terms*— FinFET, EDFinFET, LER

## I. INTRODUCTION

THE arrival of FinFETs at 22 [1] and 14 nm [2] node has continued the CMOS scaling. However in FinFETs, the requirement to have a thin fin width (Wfin ~ $L_G$/3 [3]) for an adequate electrostatic control, increases the impact of LER on the device variability and makes it one of the most critical variability component [4]. To address $V_T$ variability challenge in FinFETs, we had earlier proposed a new device architecture [5][6], – the Epitaxially Defined FinFET (EDFinFET). In EDFinFET (Fig. 1(a)), a low-doped channel is grown by conformal Si epitaxy on the highly doped Si fin. The depletion width, being defined by undoped Si epitaxy, remains uniform. This is true in spite of the LER in the heavy doped fin beneath (Fig. 1(b-ii)), as the underlying heavily doped fin width cannot be depleted. Therefore, even with LER in the starting fin, depletion width is unaffected in EDFinFET, leading to reduction in $V_T$ variability. EDFinFET in Dynamic Threshold MOS configuration (DTMOS) [5] is referred to as DTEDFinFET. Key results from [5] are reproduced in Fig. 1 (c), wherein it can be seen that both EDFinFET and DTEDFinFET reduce $V_T$ variability by 35% and 59% respectively with respect to least variable FinFET [7] ($W_{FIN}$ = 5nm). DTEDFinFET also gives 43% higher $I_{ON}$. Also shown is the challenge of FinFET scalability as further reduction in $W_{FIN}$ results in the rapid rise in $V_T$ variability.

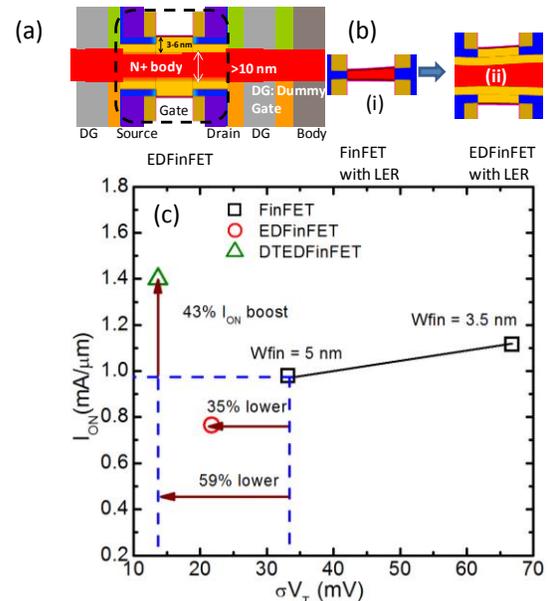

Fig. 1(a). 2D cut of EDFinFET device architecture, (b) FinFET (i) and EDFinFET (ii) subjected to same extent of LER. EDFinFET is immune to LER $V_T$ variability since channel width is defined by epitaxy (c) Comparison of σ$V_T$ with $I_{ON}$ for FinFET at $L_G$ = 15 nm, $W_{FIN}$ = 3.5 nm (LER prone) and 5 nm (less prone to LER), EDFinFET and DTEDFinFET. DTEDFinFET shows 43% $I_{ON}$ boost and 59% reduction in $V_T$ variability.

In this paper we present a circuit density comparison of EDFinFET and FinFET. Circuit density is an important metric for CMOS scaling as increased density leads to reduced cost per unit transistor. However, as discussed in [5], EDFinFET consists of wider fins and thus it may result in less dense circuits. In this paper we show by a simple formulation that even though individual EDFinFET transistor may consume larger print area, at circuit level, EDFinFET may produce denser circuits.

The paper is arranged as follows. In section II, we present the formulation of circuit density analysis. Results of the analysis are discussed in section III. Conclusions are summarized in Section IV.

This work was supported by the Dept. of Science and Technology, India. And Applied Materials Inc., Santa Clara, CA 94085.

S. Mittal, S. Lodha, and U. Ganguly are with the Department of Electrical Engineering, Indian Institute of Technology Bombay, Mumbai, India – 400 076(e-mail: smittal, udayan@ee.iitb.ac.in).

A. Nainani and M.C. Abraham are with Applied Materials Inc., Santa Clara, USA, 94085.

## II. FORMULATION

The layout of EDFinFET and FinFET is compared in Fig. 2. Wider fins and an individual body contact for EDFinFET increases the area consumption for each transistor [5]. EDFinFET consumes 3 gate pitches unlike 2 for FinFETs due to an additional body contact as shown in Fig. 2 which leads to about 2× higher transistor area. To compare current drive per unit area, the effective channel width is needed. The effective channel width in FinFET and EDFinFET is proportional to the height of fin structure. Mechanical stability concerns of fins due to surface tension during wet processing of wafer (post STI etch wet clean etc.) has been shown to limit fin $H_{FIN}$, such that maximum $H_{FIN}$ is given as [8]:

$$H_{FIN} = \frac{8 W_{FIN} S^2 E \rho}{9\gamma} \quad (i)$$

where $W_{FIN}$ is fin width as defined in Fig. 2, S is spacing between two fins, E is Young's Modulus of silicon, ρ is density of silicon, and γ is surface tension of liquid used for wet processing. Please note that for EDFinFET, S is the space between two heavy doped inner fins which experiences the wet clean step and not the space finally left to fill (S') after epitaxial deposition. Similarly $W_{FIN}$ is starting heavy doped fin's width and not the total width after epitaxial deposition. For typical numbers of E, ρ and γ from [8], equation (i) reduces to:

$$H_{FIN}\ (nm) = 0.015\ W_{FIN}(nm) S^2(nm^2) \quad (ii)$$

where $W_{FIN}$ = (P-S) (P is Pitch). Equation (ii) is a cubic equation in S. The $H_{FIN}$ thus obtained from equation (ii) for three different pitches of 50, 70 and 90 nm is plotted against spacing S and shown in Fig. 3. Maximum possible $H_{FIN}$ is desirable to get maximum effective channel width, and its relation can be obtained by differentiating equation (ii) with respect to S and is shown in equation (iii):

$$H_{FIN-MAX}\ (nm) = 4 \times 0.015 \frac{(P)^3 (nm^3)}{27} \quad (iii);$$
$$where\ S = \frac{2P}{3}$$

Thus, to obtain maximum $H_{FIN}$ at 50 nm pitch, the spacing S needs to be 33.3 nm (= 2P/3), and which corresponds to a fin width $W_{FIN}$ of 16.7 nm (=P-S). However for FinFETs, $W_{FIN}$ is constrained to ≤ 5 nm (~$L_G$/3) for adequate electrostatic control [3] and minimum $V_T$ variability [7]. So it is not possible to achieve maximum possible $H_{FIN}$ for FinFET technology because of electrostatic and variability constraints on fin width. $H_{FIN}$ for FinFET at different pitches is shown by red symbols in Fig. 3, where space is such that fin width is 5 nm. On the other hand, the 16.7 nm fin width is thickness of starting heavy doped fin ($W_{FIN}$ in Fig. 2 (b)) for EDFinFET. So it is possible to use the maximum fin height in EDFinFET as starting fin width of EDFinFET is not limited by electrostatics unlike FinFETs where fin width ≤ 5 nm. Rather, the thin epi thickness of 6 nm will ensure good electrostatic control of the EDFinFET. In terms of gap-fill concerns, with this epi thickness, actual space left to be filled by STI is 21.3 nm (S'=S-2*epi), which is a standard STI gap-fill requirement for 25 nm NAND flash and has been demonstrated in [9] and [10]. Further pitch relaxation, may increase space for ease of STI gap-fill while enabling even taller $H_{FIN}$. Thus obtained maximum $H_{FIN}$ for EDFinFET is shown in Fig. 3 by green symbols for different pitches.

Since maximum $H_{FIN}$ is possible for EDFinFET, it increases strongly for EDFinFET compared to FinFETs with the increment of pitch (Fig. 3). For comparison, EDFinFET has 3× $H_{FIN}$ compared to FinFETs at 90nm pitch, while at 50 nm pitch, it is 1.8×. However, the useful $H_{FIN}$ for current conduction is above STI. So by discounting a realistic STI depth of 60 nm [11], [12], the effective $H_{FIN}$ advantages in EDFinFET are even better. $H_{FIN}$ of this order has been demonstrated in [13].

Table I shows the effective height obtained for three devices along with current benefits for EDFinFET and DTEDFinFET. Please note that an optimized fixed pitch of 50 nm [14] is chosen for FinFET and pitch is varied for EDFinFET technology for optimization. Data for 50 and 70 nm pitch is shown in table I. Equation (iv) is the final equation which governs the height of FinFET and equation (v) shows the same for EDFinFET.

$$H_{FIN-FinFET}\ (nm) = 0.015\ W_{FIN}(P - W_{FIN})^2 - STI \quad (iv)$$
$$H_{FIN-EDFinFET}\ (nm) = H_{FIN-MAX}(eq.(iii)) - STI \quad (v);$$

TABLE I
$H_{FIN}$, $I_{ON}$/FIN, $I_{ON}$/AREA COMPARISON FOR THREE DEVICES

|  | $I_{ON}$ (mA)/μm (A) | Pitch (P) (nm) | $H_{FIN}$ effective* (B) (nm) | $I_{ON}$/fin = A× 2B (mA) | Area with respect to FinFETs | $I_{ON}$/ Area (mA) |
|---|---|---|---|---|---|---|
| FinFET | 0.98 | 50 | 92 | 0.18 | 1 | 0.18 |
| EDFinFET | 0.76 | 50 | 218 | 0.33 | 1.5 | 0.22 |
|  |  | 70 | 702 | 1.07 | 2.1 | 0.51 |
| DT-EDFinFET | 1.4 | 50 | 218 | 0.61 | 1.5 | 0.41 |
|  |  | 70 | 702 | 1.97 | 2.1 | 0.94 |

*Equation (iv) for FinFET; equation (v) for EDFinFET

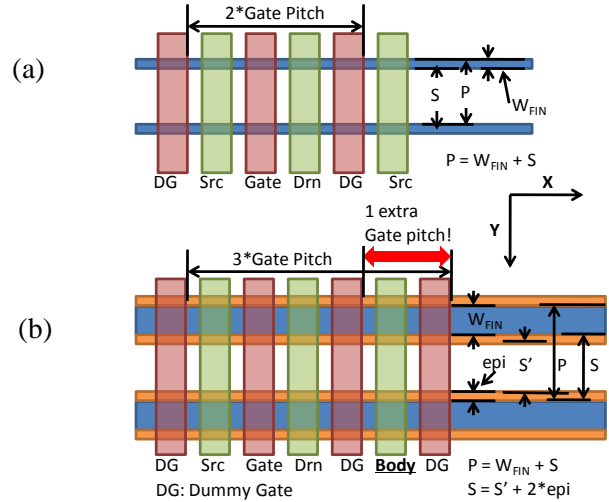

Fig. 2 (a). Layout of FinFET (assuming self-aligned contacts); only 2 gate pitches are required to make gate and self-aligned source, and drain. (b) Layout of EDFinFET; due to body contact, 3 gate pitches are required. Also showing different notations for FinFET and EDFinFET: P is pitch, $W_{FIN}$ is fin width (n+ starting fin width for EDFinFET), S is space between two fins for FinFET and distance between two starting heavy doped fins for EDFinFET, S' is final space between two EDFinFET fins after epitaxy that is then subject to STI gap-fill and epi is epitaxial layer thickness for EDFinFET.

The height benefit and the resultant single fin current benefits for EDFinFET & DTEDFinFET over FinFET are plotted in Fig. 4. With the increment in pitch as shown on the x-axis, $H_{FIN}$ of EDFinFET rise rapidly over 50 nm pitch FinFET. With 80% increment in pitch, EDFinFET can be as much as 17× taller than FinFETs as compared to 2.3× for same pitch. Also, even though $I_{ON}/\mu m$ is less for EDFinFET in comparison to FinFETs, due to possibility of growing taller fin, even at 50 nm pitch, one fin of EDFinFET gives 1.8× more current. The gain increases to 13× with 80% increment in pitch. DTEDFinFET, owing to its inherent $I_{ON}$ benefits, is even better with benefits ranging from 3.3× to 24× in the range of pitches considered.

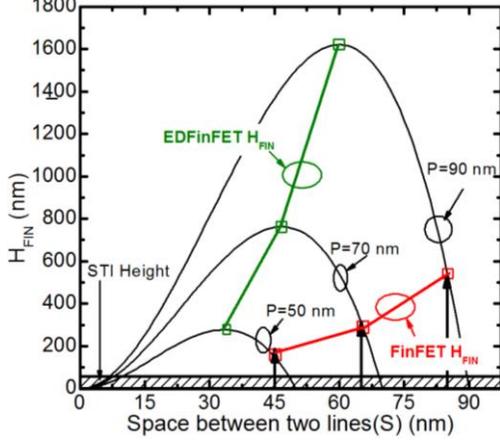

Fig. 3. $H_{FIN}$ vs. space S between two fin drawn for three different pitches of 50, 70 and 90 nm. EDFinFET is capable of achieving maximum $H_{FIN}$. Maximum $H_{FIN}$ increases with pitch (shown by green symbols for EDFinFET). FinFETs, due to narrow $W_{FIN}$ requirements based on electrostatics, cannot achieve maximum possible $H_{FIN}$. FinFET $H_{FIN}$ is shown by red symbols. The "STI height" of 60 nm i.e. the height above which would be the actual $H_{FIN}$ is marked by the hatched area.

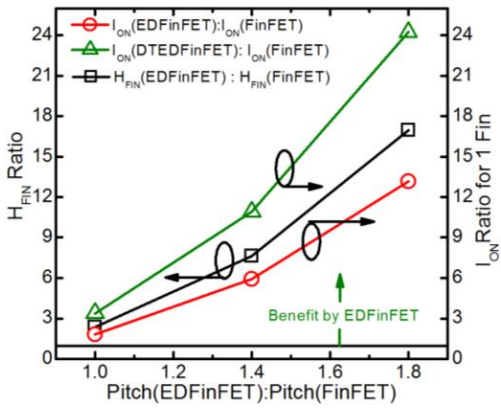

Fig. 4. $H_{FIN}$, $I_{ON}$ per fin of EDFinFET compared against FinFET for different pitch ratios at standard FinFET pitch = 50 nm [14]. EDFinFET technology can have 2 to 17 times taller fins resulting into 2 to 24 times more current per fin for EDFinFET & DTEDFinFET.

Area consumption penalty and current for same area are shown in Fig. 5 (data is shown in table I). It shows that even though a single transistor of EDFinFET consumes ~1.5-3× more layout area (depending on chosen pitch) in comparison to FinFETs, current drive per unit area can be more for EDFinFET owing to its taller fin advantage. For 50 nm pitch, EDFinFET consumes 1.5 times more area due to an extra gate pitch (Fig. 2(b)) and it still has 1.2× more current per unit area. The same advantage is 2.3× for DTEDFinFET. For 80% higher pitch, area consumed is 2.7 times more and currents per unit area for EDFinFET and DTEDFinFET are 4.9× and 9× higher respectively in comparison to FinFETs. Fig. 5 also shows the benefits of 40% increment in pitch of EDFinFETs (pitch = 70 nm). This particular pitch is referred to as "Relaxed Pitch" configuration in the discussion to follow.

At circuit level, to achieve higher $I_{ON}$, width of the channel is traditionally increased in planar technology. The well-known ratio of channel width (W) to channel length (L) (i.e. W/L) is the parameter which represents the same. In a FinFET technology, a W increase is achieved by adding additional fins or by using multiple fin heights [15]. In EDFinFET, the taller fins further enable using multiple $H_{FIN}$ by engineering different STI heights as an effective way of achieving higher (W/L) ratios. If the requirement is to go to 2 (W/L) from 1(W/L), the EDFinFET fin height can be engineered such that 2(W/L) is achieved from a single fin, instead of adding additional fins, as shown in Fig. 6 (a), (b). This may require an additional non-critical mask for selected area fin height engineering but it may enable higher circuit densities.

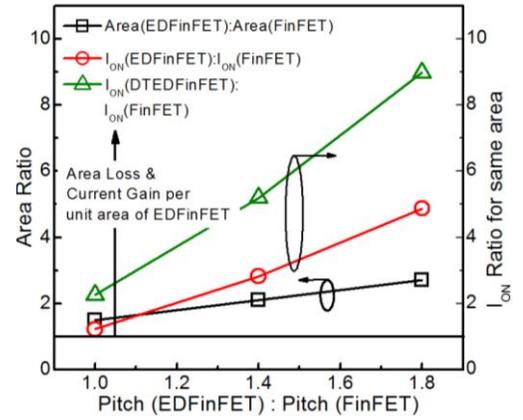

Fig. 5. Area consumed and $I_{ON}$ per unit area of EDFinFET & DTEDFinFET compared against FinFET for different pitch ratios at FinFET pitch = 50 nm [14]. Even though EDFinFET consumes more area, $I_{ON}$ per unit area can be 1.2 to 9 times more for the pitch ratio shown.

In Fig. 6(a), $H_{FIN1}$ is the maximum possible height for EDFinFET. One fin of DTEDFinFET, with maximum height $H_{FIN1}$, gives 11× higher $I_{ON}$ with respect to FinFETs in relaxed pitch configuration. By engineering various STI levels to obtain different $H_{FINS}$, up to 11(W/L)'s can be extracted from a single fin of EDFinFET without compromising performance cf. 11 fins for FinFETs. Such multiple STI heights is proposed earlier for FinFETs [14]. Thus EDFinFET technology is capable of benefits in terms of area efficiency by using multiple fin heights instead of multiple fins. Equation (vi) shows the formula to calculate number of EDFinFET fins for a CMOS inverter (Fig. 6 (c)) and equation (vii) & (viii) shows the same for a 2 input NAND Gate (Fig. 6(d)).

*For Inverter*:
$$No. of\ PMOS\ fins(P) = No. of\ NMOS\ fins\ (N)$$
$$= \left\lceil \frac{n}{I_{RATIO}} \right\rceil \quad (vi);$$

*For NAND gate*:
$$No. of\ PMOS\ fins(P) = 2 \times \left\lceil \frac{n}{I_{RATIO}} \right\rceil \quad (vii);$$

$$No.\,of\,NMOS\,fins(N) = 2 \times \left\lceil \frac{2n}{I_{RATIO}} \right\rceil \quad (viii);$$

where $\lceil\,\rceil$ is ceiling function, $n$ is minimum $\left(\frac{W}{L}\right)$ ratio in circuit, and $I_{RATIO} = \dfrac{\dfrac{I_{ON}}{fin} FinFET}{\dfrac{I_{ON}}{fin} EDFinFET\,(or\,DTEDFinFET)}$

Total number of EDFinFET fins with this methodology would be P+N. While for FinFET; number of fins equals total number of (W/L) ratio used in circuit. Therefore for inverter, number of FinFET fins equals 2n, and for 2-input NAND gate, it is 6n.

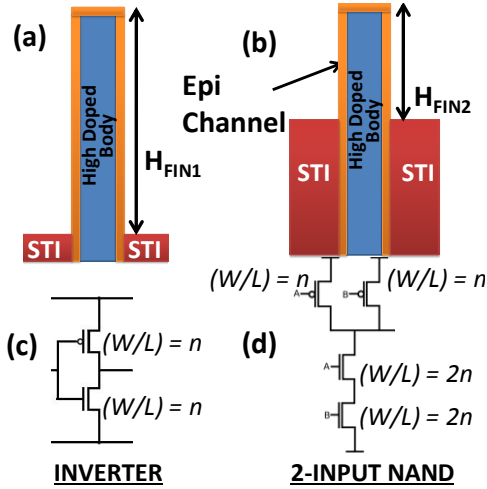

Fig. 6. Taller Fins for EDFinFET may produce different $H_{FINS}$ by engineering different STI heights. (a) Cartoon showing $H_{FIN1}$, the maximum possible height for EDFinFET, (b) cartoon showing height of fin as $H_{FIN2}$, changed by modulating the STI depth. So $H_{FIN2}$ can be used for single (W/L) transistor and $H_{FIN1}$ can be used for double (W/L) transistor, (c) Schematic of CMOS Inverter and (d) 2-input NAND Gate. (W/L) ratio is annotated by n.

## III. RESULTS AND DISCUSSIONS

The area benefit obtained with this methodology at circuit level is shown in Fig. 7 (a) for EDFinFET and Fig. 7 (b) for DTEDFinFET for a basic CMOS inverter and 2 input NAND gate, for different (W/L) ratios. Ratio of area consumed by FinFET to same pitch and relaxed pitch EDFinFET (and DTEDFinFET) configurations is plotted for different values of (W/L). (W/L) in this plot refers to minimum ratio used in the circuit concerned.

Relaxed pitch EDFinFET, due to taller fin advantage, starts producing area benefit for 2(W/L) and above with the use of above mentioned multiple fin height methodology. Same pitch EDFinFET starts to break-even at 3(W/L). Taller fins in relaxed pitch configuration can give as much as 2.5× area benefit (at 5(W/L), both for Inverter and NAND gate) for EDFinFET as can be seen from Fig. 7 (a). DTEDFinFET, on the other hand starts to give area benefit at 2(W/L) and above, both in relaxed pitch and same pitch configuration. Same pitch NAND consumes same area even at 1(W/L). As much as 3.5× area benefit for 5(W/L) NAND can be obtained at higher (W/L)'s by relaxed pitch DTEDFinFET configuration. So same pitch, due to less area consumption in first place (due to having same pitch), breaks even earlier but does not give much advantage at higher (W/L)'s and relaxed pitch breaks even later but gives more benefits at higher (W/L)'s due to the capability of taller fins. Experimentally, greater than 2 (W/L)'s are used for an inverter or SRAM cell [16] and 5-50(W/L)'s are used in NAND gate buffer [17]. Various other circuits like level-shifters also use >5 (W/L) transistors [18]. Thus EDFinFET may produce much better circuit densities than FinFETs, in such circuits which use higher (W/L), without compromising on performance.

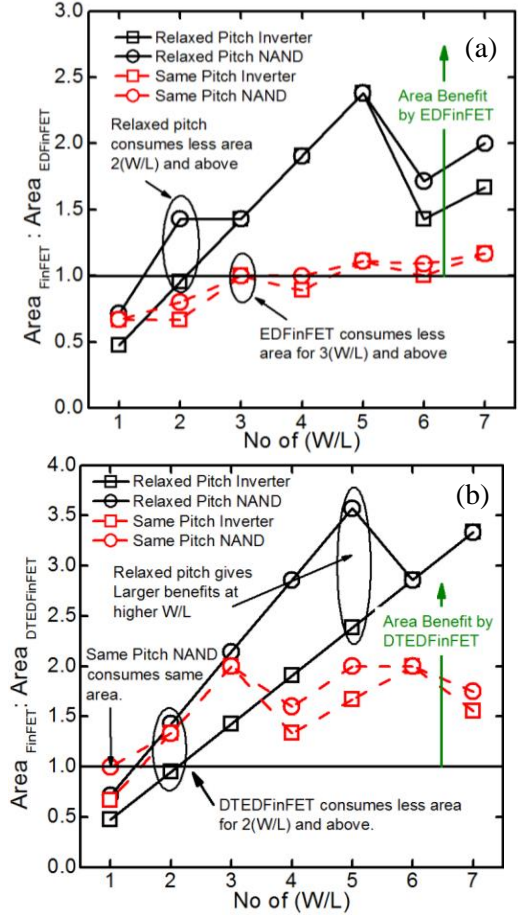

Fig. 7. Relative area cf. FinFET consumed by (a) EDFinFET and (b) DTEDFinFET vs. (W/L) ratio for inverter and 2 input NAND circuit at same pitch (50 nm) and relaxed pitch (70 nm) for two transistors. EDFinFET consumes less area for 3(W/L) s and above. DTEDFinFET takes over FinFET at 2(W/L) and above. Relaxed pitch shows larger benefits at higher (W/L).

The "jagged" nature of curve is due to the fact that increment in number of fins in EDFinFET is not simply proportional to width (W) unlike in FinFETs. $I_{ON}$ benefit for EDFinFET is around 5.9× at relaxed pitch. Therefore till 5(W/L), one fin can be used and for 6(W/L), 2 fins would be required leading to a sudden jump in area consumption and this can be confirmed by using the formula given in eq. (vi), (vii) & (viii). This jump can be observed both for EDFinFET and DTEDFinFET.

This simple discussion shows that even though one EDFinFET transistor consumes more area than a FinFET, circuit density maybe comparable or higher.

## IV. Conclusion

In this paper we discussed print area considerations for EDFinFET technology. We showed that per transistor EDFinFET consumes ~2.1× larger area because of body contact and greater pitch. However, wider fins enable taller fin heights. The ability to engineer multiple STI levels on tall fins enables different transistor widths (i.e. various W/Ls e.g. 1-10) in a single fin. This capability ensures that even though individual EDFinFET devices have ~2× larger footprints than FinFETs, EDFinFET may produce equal or higher circuit density for basic building blocks like inverters or NAND gates for W/Ls of 2 and higher, without performance penalty.

## Acknowledgment

Authors would like to thank Prof. Souvik Mahapatra and Prof. Swaroop Ganguly for discussions.

## References

[1] C. Auth, C. Allen, A. Blattner, D. Bergstrom, M. Brazier, M. Bost, M. Buehler, V. Chikarmane, T. Ghani, T. Glassman, R. Grover, W. Han, D. Hanken, M. Hattendorf, P. Hentges, R. Heussner, J. Hicks, D. Ingerly, P. Jain, S. Jaloviar, R. James, D. Jones, J. Jopling, S. Joshi, C. Kenyon, H. Liu, R. McFadden, B. McIntyre, J. Neirynck, C. Parker, L. Pipes, I. Post, S. Pradhan, M. Prince, S. Ramey, T. Reynolds, J. Roesler, J. Sandford, J. Seiple, P. Smith, C. Thomas, D. Towner, T. Troeger, C. Weber, P. Yashar, K. Zawadzki, and K. Mistry, "A 22nm high performance and low-power CMOS technology featuring fully-depleted tri-gate transistors, self-aligned contacts and high density MIM capacitors," in *2012 Symposium on VLSI Technology (VLSIT)*, 2012, pp. 131–132.

[2] S. Natarajan, M. Agostinelli, S. Akbar, M. Bost, A. Bowonder, V. Chikarmane, S. Chouksey, A. Dasgupta, K. Fischer, Q. Fu, T. Ghani, M. Giles, S. Govindaraju, R. Grover, W. Han, D. Hanken, E. Haralson, M. Haran, M. Heckscher, R. Heussner, P. Jain, R. James, R. Jhaveri, I. Jin, H. Kam, E. Karl, C. Kenyon, M. Liu, Y. Luo, R. Mehandru, S. Morarka, L. Neiberg, P. Packan, A. Paliwal, C. Parker, P. Patel, R. Patel, C. Pelto, L. Pipes, P. Plekhanov, M. Prince, S. Rajamani, J. Sandford, B. Sell, S. Sivakumar, P. Smith, B. Song, K. Tone, T. Troeger, J. Wiedemer, M. Yang, and K. Zhang, "A 14nm logic technology featuring 2nd-generation FinFET, air-gapped interconnects, self-aligned double patterning and a 0.0588 μm2 SRAM cell size," in *2014 IEEE International Electron Devices Meeting*, 2014, pp. 3.7.1–3.7.3.

[3] J. Kedzierski, M. Ieong, E. Nowak, T. S. Kanarsky, Y. Zhang, R. Roy, D. Boyd, D. Fried, and H. S. P. Wong, "Extension and source/drain design for high-performance FinFET devices," *IEEE Trans. Electron Devices*, vol. 50, no. 4, pp. 952–958, 2003.

[4] X. Wang, A. R. Brown, B. Cheng, and A. Asenov, "Statistical variability and reliability in nanoscale FinFETs," in *Technical Digest - International Electron Devices Meeting, IEDM*, 2011.

[5] S. Mittal, S. Gupta, A. Nainani, M. C. Abraham, K. Schuegraf, S. Lodha, and U. Ganguly, "Epitaxially defined FinFET: Variability resistant and high-performance technology," *IEEE Trans. Electron Devices*, vol. 61, no. 8, pp. 2711–2718, 2014.

[6] S. Mittal, S. Gupta, A. Nainani, M. Abraham, K. Schuegraf, S. Lodha, and U. Ganguly, "EDFinFET to reduce variability and enable multiple VT," in *Drc*, 2011, vol. 7, no. 4, pp. 6–7.

[7] S. Mittal, A. S. Shekhawat, and U. Ganguly, "FinFET scaling rule based On variability considerations," in *2015 73rd Annual Device Research Conference (DRC)*, 2015, pp. 127–128.

[8] C. Jungdal, "Technology and Design of TANOS-based NAND flash memory," in *International Memory Workshop*, 2010, pp. 1–36.

[9] K. Prall and K. Parat, "25nm 64Gb MLC NAND technology and scaling challenges invited paper," in *2010 International Electron Devices Meeting*, 2010, pp. 5.2.1–5.2.4.

[10] Kinam Kim, "Technology for sub-50nm DRAM and NAND flash manufacturing," in *IEEE InternationalElectron Devices Meeting, 2005. IEDM Technical Digest.*, pp. 323–326.

[11] A. Kaneko, A. Yagishita, K. Yahashi, T. Kubota, M. Omura, K. Matsuo, I. Mizushima, K. Okano, H. Kawasaki, S. Inaba, T. Izumida, T. Kanemura, N. Aoki, K. Ishimaru, H. Ishiuchi, K. Suguro, K. Eguchi, and Y. Tsunashima, "Sidewall transfer process and selective gate sidewall spacer formation technology for sub-15nm finfet with elevated source/drain extension," in *IEEE InternationalElectron Devices Meeting, 2005. IEDM Technical Digest.*, pp. 844–847.

[12] H. Kawasaki, K. Okano, A. Kaneko, A. Yagishita, T. Izumida, T. Kanemura, K. Kasai, T. Ishida, T. Sasaki, Y. Takeyama, N. Aoki, N. Ohtsuka, K. Suguro, K. Eguchi, Y. Tsunashima, S. Inaba, K. Ishimaru, and H. Ishiuchi, "Embedded Bulk FinFET SRAM Cell Technology with Planar FET Peripheral Circuit for hp32 nm Node and Beyond," in *2006 Symposium on VLSI Technology, 2006. Digest of Technical Papers.*, pp. 70–71.

[13] D. Cohen-Elias, J. J. M. Law, H. W. Chiang, A. Sivananthan, C. Zhang, B. J. Thibeault, W. J. Mitchell, S. Lee, a. D. Carter, C. Y. Huang, V. Chobpattana, S. Stemmer, S. Keller, and M. J. W. Rodwell, "Formation of sub-10 nm width InGaAs finFETs of 200 nm height by atomic layer epitaxy," in *Device Research Conference - Conference Digest, DRC*, 2013, vol. 2011, no. SUPPL., pp. 19–20.

[14] A. B. Sachid, "Technology-Circuit co-design using FinFETs at sub-22 nm nodes," IIT Bombay, 2011.

[15] A. B. Sachid and C. Hu, "Denser and More Stable SRAM Using FinFETs With Multiple Fin Heights," *IEEE Trans. Electron Devices*, vol. 59, no. 8, pp. 2037–2041, Aug. 2012.

[16] "Advanced Technologies on SRAM." [Online]. Available: http://www-inst.eecs.berkeley.edu/~ee290d/fa13/LectureNotes/Lecture14.pdf.

[17] K. Duvvada, "HIGH SPEED DIGITAL CMOS INPUT BUFFER DESIGN," Boise State University, 2006.

[18] Y. Moghe, T. Lehmann, and T. Piessens, "Nanosecond Delay Floating High Voltage Level Shifters in a 0.35 μm HV-CMOS Technology," *IEEE J. Solid-State Circuits*, vol. 46, no. 2, pp. 485–497, Feb. 2011.